\documentclass[twocolumn,showpacs,preprintnumbers,amsmath,amssymb]{revtex4}
\usepackage{graphicx}

\newcommand{\be}{\begin{equation}}
\newcommand{\ee}{\end{equation}}
\newcommand{\ba}{\begin{eqnarray}}
\newcommand{\ea}{\end{eqnarray}}
\newcommand{\bra}[1]{\langle #1 |}
\newcommand{\ket}[1]{| #1 \rangle}
\newcommand{\ds}{\displaystyle}
\newcommand{\MPl}{M_{\mathrm{Pl}}}
\newcommand{\opp}[3]{{#1}_{#2}^{\phantom{#2} #3}}
\newcommand{\ned}[3]{{#1}^{#2}_{\phantom{#2} #3}}
\newcommand{\eq}{\hspace{-1mm} & = & \hspace{-1mm}}
\newcommand{\pluss}{\hspace{-1mm} & + & \hspace{-1mm}}
\newcommand{\eqequiv}{\hspace{-1mm} & \equiv & \hspace{-1mm}}
\newcommand{\vs}{\vspace{1.5mm}}
\newcommand{\stackrelraise}[3]
  {\stackrel{\raisebox{#1}[0mm][0mm]{$\scriptstyle #2$}}{#3}}
\newcommand{\Mathematica}{Mathematica}
\newcommand{\diag}{\mathrm{diag}}
\newcommand{\sgn}{\mathrm{sgn}}

\begin{document}

\title{Higher order corrections to the Newtonian potential\\
in the Randall-Sundrum model}

\author{Petter Callin}
  \email{n.p.callin@fys.uio.no}
\author{Finn Ravndal}
  \email{finn.ravndal@fys.uio.no}
\affiliation{Department of Physics, University of Oslo, N-0316 Oslo, Norway\\}

\date{November 11, 2004}

\begin{abstract}
The general formalism for calculating the Newtonian potential in fine-tuned or
critical Randall-Sundrum braneworlds is outlined. It is based on using the full
tensor structure of the graviton propagator. This approach avoids the
brane-bending effect arising from calculating the potential for a point source.
For a single brane, this gives a clear understanding of the disputed overall
factor $4/3$ entering the correction. The result can be written on a compact
form which is evaluated to high accuracy for both short and large distances.
\end{abstract}


\maketitle

\section{Introduction}

During the recent years, there has been a lot of activity on theories with
extra dimensions. In contrast to the Kaluza-Klein picture where the radius of
the extra dimension is usually taken to be the Planck length, these new
theories allow for millimeter-sized extra dimensions as first proposed by
Arkani-Hamed \textit{et al.} \cite{ADD}, or even an infinite extra dimension as
proposed by Randall and Sundrum \cite{RandallSundrum1, RandallSundrum2}. These
models were originally constructed to solve the hierarchy problem between the
electroweak scale and the Planck scale and usually require Standard Model
fields to be trapped on a 4-dimensional surface, called a 3-brane, to avoid
conflict with experiments in particle physics where no signs of extra
dimensions have been detected so far. Since gravity per definition can
propagate through all dimensions, one important question is then whether
gravity is still effectively localized on the brane, and Newton's law
reproduced for distances larger than about a millimeter.

Starting with Randall and Sundrum \cite{RandallSundrum2}, several other authors
have calculated the Newtonian potential in the RS model, finding that $V(r)
\sim \frac{1}{r}\left[1+\Delta(r)\right]$, i.e. a dominant 4-dimensional term
together with corrections. Although everyone agrees with the general form
$\Delta \sim 1/\mu^2 r^2$ of the leading order correction (for large
distances), there seems to be an ambiguity concerning the numerical
coefficient. By looking at the scalar part of the graviton propagator, the
result $\Delta = 1/(2\mu^2 r^2)$ is obtained \cite{Kiritsis, Ghoroku}, whereas
calculating the component $h_{00}$ of the metric perturbation due to a matter
source on the brane gives the result $\Delta = 2/(3\mu^2 r^2)$
\cite{Garriga_Tanaka, Giddings, Chung, Deruelle_Dolezel}. In the latter
calculation, complications arise because the position of the brane can no
longer be fixed in the fifth dimension when introducing a matter source. The
relative factor $4/3$ between the above results is then considered to be the
net result of this brane-bending effect.

In this paper we calculate the Newtonian potential from the graviton propagator
for a critical brane (with cosmological constant $\Lambda=0$ and curvature
$k=0$), and show that the factor $4/3$ is obtained by using the full tensor
structure of the propagator instead of just the scalar part, without the need
to include any brane-bending effects since we don't have a matter source
\cite{Nojiri}. This also gives the same general expression for the correction
as obtained by Chung \textit{et al.} \cite{Chung}, but derived in more detail
here. Throughout this paper, we try to include most of the details in each
calculation, hopefully making it possible to follow without much prior
knowledge on the subject. The only parts left out are either straightforward or
easily found elsewhere in the literature.

Other works which are closely related to this paper include \cite{Kiritsis} and
\cite{Ito_indgrav}, where an induced 4-dimensional curvature term (i.e. a term
$\sim \int d^4 x \sqrt{g} R$ in the action) modifies the Newtonian potential so
that it behaves 4-dimensional even at very short distances, and
\cite{Ito_ncompact}, where $n$ compact extra dimensions are included in
addition to the large one. In \cite{Singh_Dadhich} the conditions for
localization of gravity are addressed more systematically, and in
\cite{Karch_Randall} it is shown that gravity is localized also for a
non-critical brane, meaning that localization is a local property of the model,
unaffected by the geometry of space far from the brane. The localization of
other spin fields (spin 0, 1, $\frac{1}{2}$ and $\frac{3}{2}$) is studied in
\cite{Bajc}.

This paper is organized as follows: In section \ref{cha:Einsteins_eq} we begin
by solving Einstein's equations for the 5-dimensional RS model, thus obtaining
its metric. In section \ref{cha:loc_gravity} we make a perturbation to the
metric and derive the wave equation the perturbation has to satisfy, starting
with a completely general metric and then specializing to the RS case. The wave
equation is then solved, giving us the graviton wavefunction. From the
wavefunction we obtain the graviton propagator in section
\ref{cha:Newtons_law}, and finally the Newtonian potential between two point
masses on the brane, expressed as an integral over the mass $m$ of the
gravitons. In order to obtain the correct normalization for the wavefunction,
and also the correct integration measure over the mass $m$, we use a regulator
brane at a distance $y_r$ from the physical brane, and take the limit $y_r \to
\infty$ at the end of the calculation. We also expand the potential in powers
of the distance $r$ between the point masses, obtaining the leading order
corrections $2/(3\mu^2 r^2)$ and $4/(3\pi\mu r)$ for large and short distances,
respectively, along with higher order corrections.

\section{The Randall-Sundrum metric}
\label{cha:Einsteins_eq}

We start out by assuming that the 5-dimensional metric has the form
\be
  ds^2 = n^2(t,y) dt^2 - a^2(t,y) \gamma_{ij} dx^i dx^j - dy^2,
  \label{eq:RS_metric}
\ee
where
\be
  \gamma_{ij} = \left( 1 + \frac{k}{4} \delta_{lm} x^l x^m \right)^{-2}
    \delta_{ij}
\ee
is the spatial metric. The metric is thus parametrized by two scale factors
$n(t,y)$ and $a(t,y)$, with $y$ being the coordinate in the fifth dimension,
and $y=0$ taken as the position of the brane. We use the standard convention
$n(t,y=0) = 1$, which means that $t$ is the proper time on the brane. The
energy momentum tensor is taken to be $T_{MN} = T^B_{MN} + T^b_{MN}$, where
\be
  T^B_{MN} = \diag(\rho_B n^2, p_B a^2 \gamma_{ij}, p_B)
\ee
is the contribution from the bulk space (index $B$), and
\be
  T^b_{MN} = \diag(\rho_b n^2, p_b a^2 \gamma_{ij}, 0) \delta(y)
\ee
is the contribution from the brane (index $b$). Finding the Einstein tensor and
solving Einstein's equation is straightforward and has been done by many
authors (see e.g. \cite{Binetruy, Langlois, Brevik, Brevik2}), so we simply
state the results here. First, the Einstein tensor is given by
\ba
  E_{tt} \eq 3 \left[
      \frac{\dot{a}^2}{a^2} - n^2 \left(
        \frac{a''}{a} + \frac{(a')^2}{a^2} - \frac{k}{a^2} \right)
    \right], \nonumber \\
  E_{ij} \eq a^2 \gamma_{ij} \left[
    \frac{(a')^2}{a^2} + \frac{2a'n'}{an} + \frac{2a''}{a} + \frac{n''}{n}
    \right. \nonumber \\
  & & \hspace{9mm} \left. - \frac{1}{n^2} \left(
      \frac{\dot{a}^2}{a^2} - \frac{2\dot{a}\dot{n}}{an} + \frac{2\ddot{a}}{a}
    \right) - \frac{k}{a^2} \right], \label{eq:Einsteintensor} \\
  E_{ty} \eq 3 \left( \frac{n'\dot{a}}{na} - \frac{\dot{a}'}{a} \right),
    \nonumber \\
  E_{yy} \eq 3 \left[
    \frac{(a')^2}{a^2} + \frac{a'n'}{an} - \frac{1}{n^2} \left(
      \frac{\dot{a}^2}{a^2} - \frac{\dot{a}\dot{n}}{an} + \frac{\ddot{a}}{a}
    \right) - \frac{k}{a^2} \right], \nonumber
\ea
where a dot denotes the derivative with respect to $t$, and a prime the
derivative with respect to $y$. Assuming that the bulk space only contains a
cosmological constant, i.e. $\rho_B = M^3 \Lambda_B = -p_B$, Einstein's
equation $E_{MN} = M^{-3} T_{MN}$ is solved by
\be
  \; a^2(t,y) = \! \left\{ \!\! \begin{array}{lc}
    \ds a_0^2 \left( 1 - \tfrac{1}{3} M^{-3} \rho_b |y| \right) +
      (\dot{a}_0^2 + k) y^2, & \Lambda_B = 0, \vs \\
    \ds \left[ a_0^2-\frac{3}{\Lambda_B}(\dot{a}_0^2+k) \right] \cosh 2\mu y
      \; - & \\
    \ds \hspace{4mm} \frac{\rho_b a_0^2}{6\mu M^3} \sinh 2\mu |y| +
      \frac{3}{\Lambda_B} (\dot{a}_0^2 + k), & \Lambda_B < 0, \vs \\
    \ds \left[ a_0^2-\frac{3}{\Lambda_B}(\dot{a}_0^2+k) \right] \cos 2\mu y
      \; - & \\
    \ds \hspace{4mm} \frac{\rho_b a_0^2}{6\mu M^3} \sin 2\mu |y| +
      \frac{3}{\Lambda_B} (\dot{a}_0^2 + k), & \Lambda_B > 0,
  \end{array} \right.
  \label{eq:scalefactor_general}
\ee
and $n(t,y) = \dot{a}/\dot{a}_0$. Here $M$ is the 5-dimensional Planck mass,
$\mu = \sqrt{|\Lambda_B|/6}$, and $a_0 = a(t,y=0)$ denotes the scale factor on
the brane. The latter satisfies an equation similar to Friedmann's first
equation
\be
  \frac{\dot{a}_0^2}{a_0^2} = \frac{\rho_b^2}{36 M^6}  +
    \frac{\Lambda_B}{6} - \frac{k}{a_0^2} - \frac{U}{a_0^4} \, ,
  \label{eq:Friedmann1}
\ee
where $U$ is a constant of integration, often called the dark radiation term
for obvious reasons. In addition, we have the energy conservation equation
\be
  \dot{\rho_b} + 3\frac{\dot{a}_0}{a_0}(\rho_b + p_b),
\ee
just as in standard cosmology. Note, in particular, the absolute value $|y|$ in
(\ref{eq:scalefactor_general}), which is the result of the boundary conditions
\be
  \frac{[a']}{a_0} = -\tfrac{1}{3} M^{-3} \rho_b, \hspace{6mm}
  \frac{[n']}{n_0} = \tfrac{1}{3} M^{-3} (2\rho_b + 3p_b),
\ee
imposed by the delta function in $T^b_{MN}$. Here $[a']$ means the jump
discontinuity in $a'$ across the brane, i.e. $[a'] = a'(y=0^+) - a'(y=0^-) =
2a'(0^+)$. If we write the energy density $\rho_b$ on the brane as $\rho_b =
\lambda + \rho$ where $\lambda$ is a constant, (\ref{eq:Friedmann1}) takes the
more recognizable form
\be
  \frac{\dot{a}_0^2}{a_0^2} =
    \frac{\rho}{3 \MPl^2} \left( 1 + \frac{\rho}{2\lambda} \right) +
    \frac{\Lambda}{3} - \frac{k}{a_0^2} - \frac{U}{a_0^4} \, ,
  \label{eq:Friedmann1_simple}
\ee
where we have identified the effective 4-dimensional Planck mass
\be
  \MPl = (M^{-6}\lambda/6)^{-1/2} = M^3 \sqrt{6/\lambda} \, ,
  \label{eq:4d_Planckmass}
\ee
and the effective cosmological constant
\be
  \Lambda = \frac{1}{12}M^{-6}\lambda^2 + \frac{1}{2}\Lambda_B =
    \frac{1}{2} (\MPl^{-2}\lambda + \Lambda_B)
  \label{eq:4d_cosmconst}
\ee
on the brane. The constant $\lambda$ is often called the brane tension, and is
assumed to be positive. Clearly, (\ref{eq:Friedmann1_simple}) reproduces
standard 4-dimensional cosmology when $U$ is $"$small$"$ and $\lambda$
$"$large$"$. We will not discuss more precisely what this means other than
saying that $\lambda$ has to be (much) larger than the typical densities
$\rho_\mathrm{nucl.}$ of matter and radiation during nucleosynthesis.

For the remainder of this paper, we will make the simplifying assumptions that
$\rho_b = \lambda$ (the brane contains no components other than its tension)
and that $U = 0$. Under these assumptions, the scale factor $a(t,y)$ can be
factorized as $a(t,y) = a_0(t) A(y)$, where $A(y)$ follows from
(\ref{eq:scalefactor_general}):
\be
  A(y) = \left\{ \begin{array}{lc}
    \ds 1 - \frac{\lambda}{6 M^3}|y|, & \Lambda_B = 0, \vs \\
    \ds \cosh \mu y - \frac{\lambda}{6\mu M^3} \sinh \mu |y|, &
      \Lambda_B < 0, \vs \\
    \ds \cos \mu y - \frac{\lambda}{6\mu M^3} \sin \mu |y|, &
      \Lambda_B > 0.
  \end{array} \right.
  \label{eq:scalefactor_simple}
\ee
It also follows that $n(t,y) = A(y)$, and the metric is simplified to
\be
  ds^2 = A^2(y) g_{\mu\nu}(x) dx^\mu dx^\nu - dy^2,
  \label{eq:metric_simple}
\ee
where $g_{\mu\nu}(x) = dt^2 - a_0^2(t) \gamma_{ij} dx^i dx^j$ is the standard
Friedmann-Robertson-Walker (FRW) metric on the brane.

Our main focus in this paper will be the case of a vanishing effective
cosmological constant, $\Lambda = 0$, which is often called a critical brane.
We then get $\Lambda_B = -\frac{1}{6}M^{-6}\lambda^2 < 0$, meaning that the
bulk space is $AdS_5$, and $A(y) = e^{-\mu|y|}$. If we also assume a spatially
flat universe, $k=0$, the metric becomes
\be
  ds^2 = e^{-2\mu|y|} \eta_{\mu\nu} dx^\mu dx^\nu - dy^2.
\ee
This is the case originally studied in \cite{RandallSundrum1, RandallSundrum2}.
One should note that $\Lambda_B < 0$ is in fact the only possibility compatible
with observations. Since the observed value of $\Lambda$ is very small, the
requirement that $\lambda \gg \rho_\mathrm{nucl.}$ obviously gives $\MPl^{-2}
\lambda \gg |\Lambda|$. From (\ref{eq:4d_cosmconst}) we therefore get
$\Lambda_B = 2\Lambda - \MPl^{-2} \lambda \approx -\MPl^{-2} \lambda < 0$.

\section{Localization of gravity}
\label{cha:loc_gravity}

The main goal of this paper is to calculate the correction to Newton's law of
gravitation due to the presence of the fifth dimension. Since the Newtonian
potential is essentially the low energy limit of the graviton propagator, we
must find the graviton wavefunction and in particular study its localization on
the brane. The graviton is described by the traceless transverse component
$h_{ij}$ of the spatial metric perturbation. The question of finding the
linearized wave equation satisfied by $h_{ij}$ has been addressed by many
authors \cite{Langlois, Isaacson, Ford_Parker}, but since this is an integral
part of the calculation, we include the details here.

\subsection{General linearized gravity}
\label{cha:linearized_gravity}

To begin with, we assume nothing about the background metric $g_{\mu\nu}$, and
consider the general perturbed metric
\be
  \hat{g}_{\mu\nu} = g_{\mu\nu} + h_{\mu\nu},
\ee
where also the number of dimensions is arbitrary. In the following, we will
only keep terms to the first order in $h$ and its derivatives. The inverse
metric is therefore
\be
  \hat{g}^{\mu\nu} = g^{\mu\nu} - h^{\mu\nu},
\ee
and the Christoffel symbols
\ba
  \hat{\Gamma}^\mu_{\;\alpha\beta} \eq
    \Gamma^\mu_{\;\alpha\beta} + \tfrac{1}{2}g^{\mu\nu} \left(
      h_{\nu\alpha;\beta} + h_{\nu\beta;\alpha} - h_{\alpha\beta;\nu}
    \right) \nonumber \\
  \eqequiv \Gamma^\mu_{\;\alpha\beta} + S^\mu_{\;\alpha\beta} \, .
\ea
From this, we see that the perturbation $S^\mu_{\;\alpha\beta}$ to the
Christoffel symbols is in fact a tensor. (The symbol $;$ means covariant
derivative with respect to the background metric $g_{\mu\nu}$.) Next, we find
the Riemann tensor
\ba
  \hat{R}^\mu_{\;\nu\alpha\beta} \eq R^\mu_{\;\nu\alpha\beta} +
    S^\mu_{\;\nu\beta,\alpha} - S^\mu_{\;\nu\alpha,\beta} \nonumber \\
  & & + \, \Gamma^\rho_{\;\nu\beta} S^\mu_{\;\rho\alpha} +
    \Gamma^\mu_{\;\rho\alpha} S^\rho_{\;\nu\beta} -
    \Gamma^\rho_{\;\nu\alpha} S^\mu_{\;\rho\beta} -
    \Gamma^\mu_{\;\rho\beta} S^\rho_{\;\nu\alpha} \nonumber \\
  \eq R^\mu_{\;\nu\alpha\beta} +
    S^\mu_{\;\nu\beta;\alpha} - S^\mu_{\;\nu\alpha;\beta} \, ,
\ea
and thus the Ricci tensor
\ba
  \hat{R}_{\mu\nu} \eq \hat{R}^\alpha_{\;\mu\alpha\nu} =
    R_{\mu\nu} + S^\alpha_{\;\mu\nu;\alpha} - S^\alpha_{\;\mu\alpha;\nu}
    \nonumber \\
  \eq R_{\mu\nu} + \tfrac{1}{2} \left(
    {h_{\alpha\mu;\nu}}^\alpha + {h_{\alpha\nu;\mu}}^\alpha -
    {h_{\mu\nu;\alpha}}^\alpha - h_{;\mu\nu}
  \right), \hspace{7mm}
\ea
where the last expression is obtained after a short and straightforward
calculation. Finally, we obtain the curvature scalar
\be
  \hat{R} = \hat{g}^{\mu\nu} \hat{R}_{\mu\nu} =
    R - h^{\mu\nu} R_{\mu\nu} + {h^{\mu\nu}}_{;\mu\nu} - \Box h \, ,
\ee
and the Einstein tensor
\ba
  \hat{E}_{\mu\nu} \eq
    \hat{R}_{\mu\nu} - \tfrac{1}{2}\hat{R}\hat{g}_{\mu\nu} \nonumber \\
  \eq E_{\mu\nu} + \tfrac{1}{2} \left(
    {h_{\alpha\mu;\nu}}^\alpha + {h_{\alpha\nu;\mu}}^\alpha -
    {h_{\mu\nu;\alpha}}^\alpha - h_{;\mu\nu}
  \right) \nonumber \\
  & & - \tfrac{1}{2} R h_{\mu\nu} + \tfrac{1}{2} g_{\mu\nu} \left(
    h^{\alpha\beta}R_{\alpha\beta} + \Box h - {h^{\alpha\beta}}_{;\alpha\beta}
  \right). \hspace{7mm}
\ea
This can be somewhat simplified by using the trace-reversed components
$\bar{h}_{\mu\nu} \equiv h_{\mu\nu} - \tfrac{1}{2} g_{\mu\nu} h$, yielding
\ba
  \delta E_{\mu\nu} \eqequiv \hat{E}_{\mu\nu} - E_{\mu\nu} \nonumber \\
  \eq \tfrac{1}{2} (
    \opp{\bar h}{\alpha\mu;\nu}{\alpha} +
    \opp{\bar h}{\alpha\nu;\mu}{\alpha} - \nabla_\alpha^2 \bar{h}_{\mu\nu} -
    R \bar{h}_{\mu\nu} \nonumber \\
  & & \hspace{2mm} + g_{\mu\nu} \bar{h}^{\alpha\beta} R_{\alpha\beta} -
    g_{\mu\nu} \ned{\bar h}{\alpha\beta}{;\alpha\beta} ) \, .
  \label{eq:Einstein_pert}
\ea
Throughout this paper $\Box$ always means the scalar d'Alembertian, i.e. $\Box
\equiv (1/\sqrt{g}) \partial_\mu (\sqrt{g} g^{\mu\nu} \partial_\nu)$, whereas
$\nabla_\alpha^2 \equiv g^{\alpha\beta} \nabla_\alpha \nabla_\beta$ means the
covariant Laplacian which acts differently on tensors of different rank. It is
therefore only for a scalar field $\phi$ that $\nabla_\alpha^2 \phi = \Box
\phi$.

The invarianse of (\ref{eq:Einstein_pert}) under an arbitrary coordinate
transformation can be used to simplify the expression. We choose to work in the
gauge where $\bar{h} = 0$ and $\ned{\bar h}{\mu\nu}{;\nu} = 0$. In order to
utilize the latter condition in (\ref{eq:Einstein_pert}), we use the general
tensor identity
\be
  T_{\alpha\beta;\mu\nu} - T_{\alpha\beta;\nu\mu} =
    T_{\alpha\sigma} R^\sigma_{\;\beta\mu\nu} +
    T_{\sigma\beta} R^\sigma_{\;\alpha\mu\nu} \, ,
\ee
which can be proven by $"$brute force$"$ by writing out the covariant
derivatives explicitely. From this it follows that
\be
  \opp{\bar h}{\alpha\mu;\nu}{\alpha} =
    \ned{\bar{h}}{\alpha}{\mu;\alpha\nu} +
    \ned{\bar{h}}{\alpha}{\sigma} R^\sigma_{\;\mu\nu\alpha} +
    \ned{\bar{h}}{\sigma}{\mu} R_{\sigma\nu} \, ,
\ee
and (\ref{eq:Einstein_pert}) is then reduced to
\ba
  \delta E_{\mu\nu} \eq
    \ned{\bar{h}}{\alpha}{\sigma} R^\sigma_{\;\mu\nu\alpha} +
    \tfrac{1}{2} \left(
      \ned{\bar{h}}{\sigma}{\mu} R_{\sigma\nu} +
      \ned{\bar{h}}{\sigma}{\nu} R_{\sigma\mu} \right. \nonumber \\
  & & \left. - \nabla_\alpha^2 \bar{h}_{\mu\nu} -
      R \bar{h}_{\mu\nu} + g_{\mu\nu} \bar{h}^{\alpha\beta} R_{\alpha\beta}
    \right).
  \label{eq:Einstein_pert2}
\ea
Taking the trace, the first three terms are seen to cancel, yielding
\be
  \delta {E^\mu}_\mu = \frac{1}{2} \! \left(
    - \Box \bar{h} - R \bar{h} + D \bar{h}^{\alpha\beta} R_{\alpha\beta}
  \right) = \frac{D}{2} \bar{h}^{\alpha\beta} R_{\alpha\beta}, \!\!
  \label{eq:Einstein_trace}
\ee
where $D$ is the number of spacetime dimensions, and we have used the traceless
condition $\bar{h} = 0$. From now on, we can also ignore the bar, since
$\bar{h}_{\mu\nu} = h_{\mu\nu}$.

Einstein's equation for the perturbed metric can be written $\hat{E}_{\mu\nu} =
M_D^{2-D}\,\hat{T}_{\mu\nu}$, where $M_D$ is the $D$-dimensional Planck mass,
and $\hat{T}_{\mu\nu}$ the perturbed energy momentum tensor. Assuming that both
energy density, pressure and $"D$-velocity$"$ $u^\mu$ is unperturbed, i.e.
$\delta\rho = \delta p = 0$ and $\delta u^\mu = 0$, we get
\ba
  \hat{T}_{\mu\nu} \eq (\rho + p) u_\mu u_\nu - p \hat{g}_{\mu\nu} \nonumber \\
  \eq (\rho + p) u_\mu u_\nu - p ( g_{\mu\nu} + h_{\mu\nu} ) \, ,
\ea
meaning that $\delta T_{\mu\nu} = -p h_{\mu\nu}$. The background metric is
determined from the unperturbed energy momentum tensor, i.e. $E_{\mu\nu} =
M_D^{2-D}\, T_{\mu\nu}$, and the equation for the perturbed Einstein tensor is
therefore
\be
  \delta E_{\mu\nu} =
    M_D^{2-D}\, \delta T_{\mu\nu} = -M_D^{2-D} p h_{\mu\nu} \, .
\ee
From (\ref{eq:Einstein_trace}) we then get
\be
  \delta {E^\mu}_\mu = \frac{D}{2} h^{\alpha\beta} R_{\alpha\beta} =
    -M_D^{2-D} p {h^\mu}_\mu = 0,
\ee
since $h$ is traceless. When inserted into (\ref{eq:Einstein_pert2}), this
gives
\ba
  2 {h^\alpha}_\sigma R^\sigma_{\;\mu\nu\alpha} +
    {h^\sigma}_\mu R_{\sigma\nu} + {h^\sigma}_\nu R_{\sigma\mu} & &
    \nonumber \\
  - \nabla_\alpha^2 h_{\mu\nu} - R h_{\mu\nu} +
    2 M_D^{2-D} p h_{\mu\nu} \eq 0 \, .
  \label{eq:Einstein_pert3}
\ea
The first four terms are often called the \textit{de Rham-Lichnerowicz
operator} $\triangle$ applied to $h_{\mu\nu}$:
\be
  \triangle h_{\mu\nu} \equiv 2 {h^\alpha}_\sigma R^\sigma_{\;\mu\nu\alpha} +
    {h^\sigma}_\mu R_{\sigma\nu} + {h^\sigma}_\nu R_{\sigma\mu} -
    \nabla_\alpha^2 h_{\mu\nu} \, .
\ee
The general equation (\ref{eq:Einstein_pert3}) can be applied to any geometry
we are interested in for tensor fluctuations.
\[ \]

\subsection{Linearized gravity in the RS model}
\label{cha:linearized_gravity_RS}

We will now apply (\ref{eq:Einstein_pert3}) to the RS background metric
(\ref{eq:RS_metric}). In addition to ${h^{MN}}_{;N} = 0$ and ${h^M}_M = 0$, we
choose the gauge where $h_{\mu 4} = h_{44} = 0$. Having only one brane, this
was shown possible in a more detailed discussion about gauge transformations by
Boos \textit{et al.} \cite{Boos}. The condition that $h_{44} = 0$ corresponds
to having no massless scalar field (radion) in the problem. (With two branes
such a gauge choice is not possible, and the radion will contribute in that
case.)

We now have a total of 10 gauge conditions, reducing the number of independent
components of the symmetric tensor $h_{MN}$ from 15 to 5. These components must
then be separated into scalar, vector and tensor parts with respect to the
spatial metric $\gamma_{ij}$. Here we choose to work only with the tensor part
which describes pure gravitational waves, meaning that we set $h_{\mu 0} = 0$.
The remaining part $h_{ij}$ therefore have two independent degrees of freedom.
If we also factorize out $a^2$ by letting $h_{ij} \to a^2 h_{ij}$, the full
perturbed metric can be written
\be
  ds^2 = n^2(t,y) dt^2 - a^2(t,y) (\gamma_{ij} + h_{ij}) dx^i dx^j - dy^2.
\ee
The indices of $h_{ij}$ are now raised and lowered by $\gamma_{ij}$, and the
gauge conditions ${h^{MN}}_{;N} = 0$ and ${h^M}_M = 0$ are reduced to $\nabla_j
h^{ij} = 0$ and ${h^i}_i = \gamma^{ij} h_{ij} = 0$, where $\nabla_j$ is the
spatial covariant derivative. Thus $h_{ij}$ may be regarded effectively as a
3-dimensional tensor, with $t$ and $y$ as additional parameters. The different
terms in (\ref{eq:Einstein_pert3}) are then easily found using {\Mathematica}
\cite{Mathematica}, with the result
\begin{widetext} \vspace{-4mm}
\ba
  2 h^K_{\;L} R^L_{\;MNK} + h^K_{\;M} R_{KN} \pluss h^K_{\;N} R_{KM} =
  \left[
    -\frac{6k}{a^2} + \frac{6(a')^2}{a^2} + \frac{2a'n'}{an} +
    \frac{2a''}{a} - \frac{6\dot{a}^2}{a^2 n^2} +
    \frac{2\dot{a}\dot{n}}{an^3} - \frac{2\ddot{a}}{an^2}
  \right] h_{MN} \, , \nonumber \\
  R = R^M_{\;M} \eq -\frac{6k}{a^2} + \frac{6(a')^2}{a^2} + \frac{6a'n'}{an} +
    \frac{6a''}{a} + \frac{2n''}{n} - \frac{6\dot{a}^2}{a^2 n^2} +
    \frac{6\dot{a}\dot{n}}{an^3} - \frac{6\ddot{a}}{an^2} \, , \nonumber \\
  2 M^{-3} p \eq -\frac{2k}{a^2} + \frac{2(a')^2}{a^2} + \frac{4a'n'}{an} +
    \frac{4a''}{a} + \frac{2n''}{n} - \frac{2\dot{a}^2}{a^2 n^2} +
    \frac{4\dot{a}\dot{n}}{an^3} - \frac{4\ddot{a}}{an^2} \, .
\ea
\end{widetext}
The last equation also follows directly from the Einstein tensor
(\ref{eq:Einsteintensor}), since $E_{ij} = M^{-3} T_{ij} = M^{-3} p a^2
\gamma_{ij}$, with $p = p_B + p_b \delta(y)$. Adding everything together, we
see that most of the terms cancel, and we are left with
\be
  \frac{1}{a^2} \nabla_K^2 (a^2 h_{ij}) = \left[
    -\frac{2k}{a^2} + \frac{2(a')^2}{a^2} - \frac{2\dot{a}^2}{a^2 n^2}
  \right] h_{ij} \, .
\ee
A tedious but straightforward calculation shows that the 5-dimensional
Laplacian can be expanded as
\ba
  & & \hspace{-6mm} \frac{1}{a^2} \nabla_K^2 (a^2 h_{ij}) = \nonumber \\
  & & \left[
    \frac{1}{n^2} \partial_0^2 +
    \left( \frac{3\dot{a}}{an^2} - \frac{\dot{n}}{n^3} \right) \partial_0 -
    \frac{1}{a^2} \gamma^{kl} \nabla_k \nabla_l - \partial_y^2
  \right. \nonumber \\
  & & \left. \;\: -
    \left( \frac{3a'}{a} + \frac{n'}{n} \right) \partial_y +
    \frac{2(a')^2}{a^2} - \frac{2\dot{a}^2}{a^2 n^2}
  \right] h_{ij} \, ,
\ea
with the result
\ba
  & & \hspace{-8mm} \left[
    \frac{1}{n^2} \partial_0^2 +
    \left( \frac{3\dot{a}}{an^2} - \frac{\dot{n}}{n^3} \right) \partial_0 -
    \frac{1}{a^2} \gamma^{kl} \nabla_k \nabla_l
  \right. \nonumber \\
  & & \left. -
    \partial_y^2 -
    \left( \frac{3a'}{a} + \frac{n'}{n} \right) \partial_y +
    \frac{2k}{a^2}
  \right] h_{ij} = 0 \, .
  \label{eq:graviton_waveeq_general}
\ea
(Removing the $y$-dependence and setting $n=1$, we recover the result for the
4-dimensional FRW metric as obtained by Ford and Parker \cite{Ford_Parker}.) We
should compare (\ref{eq:graviton_waveeq_general}) to the wave equation of a
scalar field with mass $m_\phi$ in five dimensions:
\ba
  & & \hspace{-6mm} (\Box + m_\phi^2) \phi = \nonumber \\
  & & \left[
    \frac{1}{n^2} \partial_0^2 +
    \left( \frac{3\dot{a}}{an^2} - \frac{\dot{n}}{n^3} \right) \partial_0 -
    \frac{1}{a^2} \gamma^{kl} \nabla_k \nabla_l
  \right. \nonumber \\
  & & \left. \;\: -
    \partial_y^2 -
    \left( \frac{3a'}{a} + \frac{n'}{n} \right) \partial_y + m_\phi^2
  \right] \phi = 0 \, .
\ea
The only difference between the two expressions lies in the tensor structure of
the term $\gamma^{kl} \nabla_k \nabla_l h_{ij}$. The graviton field has also
acquired an effective mass $m_k^2 = 2k/a^2$, which at the present epoch is of
the order $10^{-33}$ eV (at most), and therefore totally negligible at short
(i.e. non-cosmological) distances. For a spatially flat universe ($k=0$),
(\ref{eq:graviton_waveeq_general}) is reduced to $\Box h_{ij} = 0$. The
components of $h_{ij}$ independently satisfy the Klein-Gordon equation in this
case, and the graviton field is thus equivalent to a set of two independent
scalar fields.

From (\ref{eq:graviton_waveeq_general}) we see that the $\mathbf{x}$-dependence
of $h_{ij}$ can be separated from the $t$- and $y$-dependence. Writing $h_{ij}
= \psi(t,y) G_{ij}(\mathbf{x})$, (\ref{eq:graviton_waveeq_general}) is solved
by $\gamma^{kl} \nabla_k \nabla_l G_{ij} = -\sigma^2 G_{ij}$ and
\ba
  & & \left[
    \frac{1}{n^2} \partial_0^2 +
    \left( \frac{3\dot{a}}{an^2} - \frac{\dot{n}}{n^3} \right) \partial_0 -
    \partial_y^2 \right. \nonumber \\
  & & \left. \hspace{2mm}
    - \left( \frac{3a'}{a} + \frac{n'}{n} \right) \partial_y +
    \frac{2k+\sigma^2}{a^2}
  \right] \psi(t,y) = 0 \, . \hspace{5mm}
\ea
The eigenvalues $\sigma^2$ of the 3-dimensional Laplacian can be found in
\cite{Ford_Parker}, but we will not need them here. Instead, since we are
considering the factorized metric (\ref{eq:metric_simple}), it is more
convenient to separate the $y$-dependence only.

Substituting $a(t,y) = a_0(t) A(y)$ and $n(t,y) = A(y)$ into
(\ref{eq:graviton_waveeq_general}), it simplifies to
\ba
  \hspace{-10mm} & & \left[
    \partial_y^2 + \frac{4A'}{A}\partial_y - \frac{1}{A^2} \right.
    \nonumber \\
  \hspace{-10mm} & & \left. \hspace{2mm} \times \! \left(
      \partial_0^2 + \frac{3\dot{a}_0}{a_0}\partial_0 -
      \frac{1}{a_0^2} \gamma^{kl} \nabla_k \nabla_l + \frac{2k}{a_0^2}
    \right)
  \right] h_{ij} = 0 \, ,
  \label{eq:graviton_5d}
\ea
with the solution $h_{ij}(x,y) = G_{ij}(x) \Phi(y)$, where
\ba
  & \ds \hspace{-7mm} \left( \!
    \partial_0^2 + \frac{3\dot{a}_0}{a_0}\partial_0 -
    \frac{1}{a_0^2} \gamma^{kl} \nabla_k \nabla_l + \frac{2k}{a_0^2} + m^2
  \! \right) \! G_{ij}(x) = 0 , &
    \label{eq:graviton_4dpart} \\
  & \ds \Phi''(y) + \frac{4A'}{A} \Phi'(y) + \frac{m^2}{A^2} \Phi(y) = 0 , &
    \label{eq:graviton_5dpart}
\ea
and $m$ is the eigenvalue following from the separation of variables. This is
the same result as obtained by Brevik \textit{et al.} \cite{Brevik}. Explicit
solutions to (\ref{eq:graviton_4dpart}) can be found by separating the
$t$-dependence, but we will not need them in the following. The mass spectrum
will be determined from (\ref{eq:graviton_5dpart}).

One should also note that the result (\ref{eq:graviton_5dpart}) can be obtained
without demanding that $h_{\mu 0} = 0$ when the 5-dimensional metric has the
simple form (\ref{eq:metric_simple}). Instead, the perturbed metric can then be
written
\be
  ds^2 = A^2(y) \left(g_{\mu\nu}+h_{\mu\nu}\right) dx^\mu dx^\nu - dy^2 \, ,
\ee
with the gauge conditions ${h^{\mu\nu}}_{;\nu} = 0$ and $h = g^{\mu\nu}
h_{\mu\nu} = 0$, where the indices of $h_{\mu\nu}$ are raised and lowered by
$g_{\mu\nu}$. Thus $h_{\mu\nu}$ is effectively a 4-dimensional tensor, with
five independent degrees of freedom. Going through the same calculations as
above, starting from (\ref{eq:Einstein_pert3}), we then get the result
\be
  \left[
    \partial_y^2 + \frac{4A'}{A} \partial_y -
    \frac{1}{A^2} \left( \nabla_\alpha^2 + \frac{2\Lambda}{3} \right)
  \right] h_{\mu\nu} = 0 \, ,
  \label{eq:graviton_5d_generell}
\ee
where $\nabla_\alpha^2$ is the 4-dimensional Laplacian. The apparent difference
in the effective mass term compared to (\ref{eq:graviton_5d}) is simply because
now we have not factorized out $a_0^2(t)$ from the perturbation. Writing
$h_{\mu\nu} = a_0^2(t) \tilde{h}_{\mu\nu}$ and using Friedmann's equation
(\ref{eq:Friedmann1_simple}), $\tilde{h}_{\mu\nu}$ can be seen to satisfy
(\ref{eq:graviton_5d}), provided we also set $\tilde{h}_{\mu 0} = 0$ in order
to expand $\nabla_\alpha^2$. Equation (\ref{eq:graviton_5d_generell}) is solved
by setting $h_{\mu\nu}(x,y) = G_{\mu\nu}(x) \Phi(y)$, where $\Phi(y)$ satisfies
(\ref{eq:graviton_5dpart}), and
\be
  \left( \nabla_\alpha^2 + \frac{2\Lambda}{3} + m^2 \right)
    G_{\mu\nu}(x) = 0 \, .
\ee
Thus, $m$ can be regarded as the effective mass of the 5-dimensional graviton
field as observed in four dimensions, with a small correction from the
cosmological constant. When $\Lambda > 0$, even the $m=0$ state is massive.
This appears to create a serious problem, since massive gravitons (in 4
dimensions) have five polarization states whereas massless gravitons only have
two polarization states, and all physical results should be continuous in the
limit $\Lambda \to 0$. This problem and its solution has been discussed by
Karch and Randall \cite{Karch_Randall}.

\subsection{The graviton wavefunction}
\label{cha:graviton_wavefunction}

Next, we proceed to solve (\ref{eq:graviton_5dpart}). Since this has already
been done \cite{RandallSundrum2, Ghoroku, Brevik}, we will only sketch the
derivation. First, we simplify the equation by making the substitution $y \to
z(y)$, where
\be
  \frac{\partial y}{\partial z} = A(y),
\ee
and by writing $\Phi(y)$ as
\be
  \Phi(y) = A^{-3/2} u(z).
\ee
This leads to the Schr{\"o}dinger like equation for $u(z)$
\be
  \left[ -\partial_z^2 + V(z) \right] u(z) = m^2 u(z),
  \label{eq:u_Schrodinger}
\ee
where the effective potential $V(z)$ is given by
\be
  V(z) = \frac{9}{4} (A')^2 + \frac{3}{2} AA''.
\ee
Notice that the line element (\ref{eq:metric_simple}) takes the form
\be
  ds^2 = A^2(z) \left( g_{\mu\nu} dx^\mu dx^\nu - dz^2 \right)
\ee
when expressed using the coordinate $z$. Therefore $z$ is often called the
conformal coordinate.

Focusing on the critical case $\Lambda = 0$ where $A(y) = e^{-\mu|y|}$, we get
the result
\ba
  z \eq \sgn(y) \frac{1}{\mu} \left( e^{\mu|y|} - 1 \right), \nonumber \\
  V(z) \eq \frac{15\mu^2}{4\left(1 + \mu|z|\right)^2} - 3\mu \delta(z) \, ,
  \label{eq:potential}
\ea
where the delta function originates from the double derivative of $|y|$, and we
have used the general relation $\delta(z-z') = A(y') \delta(y-y')$ together
with $A(0) = 1$. The potential for the case $\Lambda > 0$ can be found in e.g.
\cite{Brevik, Ito}, and in \cite{Karch_Randall} the case $\Lambda < 0$ is also
included. The delta function in the result means that $u(z)$ must depend on the
absolute value $|z|$ (just as the delta function in $T_{MN}$ resulted in
$a(t,y)$ depending on $|y|$). Matching the delta function in $V(z)$ with the
delta function from $u''(z)$, we get the boundary condition
\be
  2u'(0) + 3\mu u(0) = 0 \, ,
  \label{eq:u_boundary}
\ee
and the equation for $z > 0$
\be
  u''(z) + \left[ m^2 - \frac{15\mu^2}{4(1 + \mu z)^2} \right] u(z) = 0 \, .
  \label{eq:u_difflign}
\ee
The general solution to (\ref{eq:u_difflign}) is easily shown to be $u(z) =
\sqrt{1+\mu z} \left\{ A J_2[\tfrac{m}{\mu}(1+\mu z)] + B
Y_2[\tfrac{m}{\mu}(1+\mu z)] \right\}$ for arbitrary constants $A$ and $B$,
where $J_2(x)$ and $Y_2(x)$ are the Bessel functions of order two of the first
and second kind, respectively. In order to satisfy the boundary condition
(\ref{eq:u_boundary}), we must choose the linear combination
\ba
  u(z) \eq N_m \sqrt{1+\mu|z|} \nonumber \\
  & & \hspace{-8mm} \times \! \left\{
    Y_2[\tfrac{m}{\mu}(1+\mu|z|)] -
    \frac{Y_1(\frac{m}{\mu})}{J_1(\frac{m}{\mu})} J_2[\tfrac{m}{\mu}(1+\mu|z|)]
  \right\}, \hspace{7mm}
  \label{eq:u_solution}
\ea
where $N_m$ is a normalization constant (to be found later). This expression is
valid for all $m$, so we have a continuous mass spectrum for $m > 0$. The
solution for $m=0$ can be found either directly from (\ref{eq:u_difflign}) or
by taking the limit $m \to 0$ in (\ref{eq:u_solution}), and is given by
\be
  u_0(z) = N_0 (1+\mu|z|)^{-3/2}.
  \label{eq:u_solution_zero}
\ee
Thus gravity is localized on the brane, since the massless, bound state
$u_0(z)$ can be normalized, and has a sharp peak at $z=0$. As we will see
later, the massive states (\ref{eq:u_solution}) give the deviation from
Newton's law. Solutions for the case $\Lambda > 0$ are given in \cite{Ghoroku,
Brevik, Ito}, and for $\Lambda < 0$ in \cite{Karch_Randall}.

\section{Corrections to Newton's law}
\label{cha:Newtons_law}

Having found the graviton wavefunction, we will use this result to calculate
the static gravitational potential between two point masses both located on the
brane. As is well known from field theory, the (Fourier transformed) potential
is obtained by looking at a scattering diagram where two particles interact
through the exchange of a virtual graviton, in the limit where the energy of
the graviton goes to zero:
\be
  V(\mathbf{k}) = \lim_{k^0 \to 0}
    \raisebox{-10mm}[10mm][9mm]{
      \includegraphics[bb= 269 717 354 779]{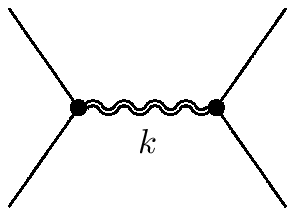}
    } .
\ee
We therefore need to know the 5-dimensional graviton propagator, as well as the
interaction between a graviton and a massive particle on the brane. The
propagator must in principle be found by inverting the quadratic terms in the
effective graviton Lagrangian \cite{Boos}. However, since we have expressed the
5-dimensional graviton wavefunction directly in terms of the 4-dimensional ones
through $h_{\mu\nu}(x,y) = \sum_m G^m_{\mu\nu}(x) \Phi(m,y)$, we know that the
5-dimensional propagator must be similarily expressed by the 4-dimensional
propagator, and can therefore immediately write down the answer:
\ba
  D^{(5)}_{\mu\nu\alpha\beta}(x,y;x',y') \eqequiv
    \bra{0} T \hat{h}_{\mu\nu}(x,y) \hat{h}_{\alpha\beta}(x',y') \ket{0}
    \nonumber \\
  \eq \sum_m \Phi(m,y) \Phi^*(m,y') D^{(4,m)}_{\mu\nu\alpha\beta}(x,x') \, ,
    \nonumber \\
  & & \label{eq:5dpropagator}
\ea
where $D^{(4,m)}_{\mu\nu\alpha\beta}(x,x')$ is the propagator of a
4-dimensional spin-2 particle with mass $m$. Taking $y=y'=0$ and using
$\Phi(m,y) = A(y)^{-3/2} u(m,z)$, we get
\be
  D^{(5)}_{\mu\nu\alpha\beta}(x,0;x',0) =
    \sum_m |u(m,0)|^2 D^{(4,m)}_{\mu\nu\alpha\beta}(x,x') \, .
  \label{eq:5dpropagator_brane}
\ee
Everything is now happening in just four dimensions. The only trace left of the
motion of the graviton through the fifth dimension is that we have a tower of
massive 4-dimensional gravitons with a non-trivial $"$weight$"$ $|u(m,0)|^2$.
More importantly, the propagator $D^{(4,m)}_{\mu\nu\alpha\beta}(x,x')$ is
completely described in the 4-dimensional FRW space with metric $g_{\mu\nu}$,
where standard results can be applied directly without worrying about the fifth
dimension.

Focusing again on the critical case $k=0$ and $\Lambda=0$, the 4-dimensional
space is flat ($g_{\mu\nu} = \eta_{\mu\nu}$), with the result
\be
  D^{(4,m)}_{\mu\nu\alpha\beta}(x,x') =
    \int \frac{d^4 k}{(2\pi)^4}
    \frac{P^{(m)}_{\mu\nu\alpha\beta}(k)}{k^2 - m^2 + i\epsilon}
    e^{-ik \cdot (x-x')} \, ,
  \label{eq:4dpropagator}
\ee
where the polarization tensor can be chosen as (see e.g. \cite{Giudice} for
more details)
\be
  P^{(m=0)}_{\mu\nu\alpha\beta}(k) = \frac{1}{2}\left(
    \eta_{\mu\alpha}\eta_{\nu\beta} + \eta_{\mu\beta}\eta_{\nu\alpha} -
    \eta_{\mu\nu}\eta_{\alpha\beta}
  \right),
  \label{eq:polarization_tensor1}
\ee \vspace{-5mm}
\ba
  & & \hspace{-2mm} P^{(m>0)}_{\mu\nu\alpha\beta}(k) =
    \frac{1}{2}\left(
      \eta_{\mu\alpha}\eta_{\nu\beta} + \eta_{\mu\beta}\eta_{\nu\alpha} -
      \eta_{\mu\nu}\eta_{\alpha\beta}
    \right) \nonumber \\
  & & \hspace{3mm} -
    \frac{1}{2m^2} \left(
      \eta_{\mu\alpha} k_\nu k_\beta + \eta_{\mu\beta} k_\nu k_\alpha +
      \eta_{\nu\alpha} k_\mu k_\beta + \eta_{\nu\beta} k_\mu k_\alpha
    \right) \nonumber \\
  & & \hspace{3mm} + \frac{1}{6}
    \left( \eta_{\mu\nu} + \frac{2}{m^2} k_\mu k_\nu \right)
    \left( \eta_{\alpha\beta} + \frac{2}{m^2} k_\alpha k_\beta \right).
  \label{eq:polarization_tensor2}
\ea

Next, we consider the interaction. For a Lagrangian ${\cal L}$ in $D$
dimensions the action is $S = \int d^D x \sqrt{g} {\cal L}$. The energy
momentum tensor $T_{MN}$ corresponding to the Lagrangian ${\cal L}$ is defined
by varying the action with respect to the metric:
\be
  \delta S \equiv \int d^D x \sqrt{g} \frac{1}{2} T_{MN} \delta g^{MN} \, .
\ee
Since $\delta g^{MN} = -h^{MN}$ to the lowest order in $h$, we therefore get
the interaction Lagrangian
\be
  {\cal L}_\mathrm{int} = -\frac{1}{2} T_{MN} h^{MN} \, .
  \label{eq:graviton_int}
\ee
So far, we have considered the metric perturbation $h_{MN}$ to be
dimensionless, since $g_{MN}$ is dimensionless. But when treating it as a field
of particles (i.e. gravitons), this can no longer be the case if we want to
have a canonical kinetic term in the Lagrangian. Instead, $h_{MN}$ has to have
a dimension $\dim[h] = (D-2)/2$ in $D$ spacetime dimensions. Since $h_{MN}$ is
a 5-dimensional field, we must therefore have $\dim[h] = 3/2$. From
$h_{\mu\nu}(x,y) = G_{\mu\nu}(x) \Phi(y)$, we then get $\dim[\Phi] = \dim[u] =
1/2$, since $G_{\mu\nu}$ is a 4-dimensional field and should therefore have
dimension 1. This is also required in order to make equations
(\ref{eq:5dpropagator}-\ref{eq:polarization_tensor2}) consistent. Again, since
$h_{\mu\nu}$ is 5-dimensional, we should use the 5-dimensional Planck mass $M$
and let $h^{\mu\nu} \to M^{-3/2} h^{\mu\nu}$ in (\ref{eq:graviton_int}).
Finally, using $M^3 = \mu \MPl^2$ which follows from (\ref{eq:4d_Planckmass})
and (\ref{eq:4d_cosmconst}), we get the result
\be
  \raisebox{-10.5mm}[10mm][9mm]{
    \includegraphics[bb= 269 717 352 779]{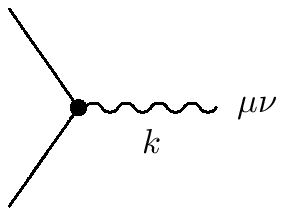}
  } \!\! = \frac{1}{\sqrt{\mu} \MPl} T^{\mu\nu}(k) \, .
  \label{eq:gravitonvertex}
\ee

Before proceeding, lets give a little thought to the assumption $\Lambda=0$ and
$k=0$. For nonzero $\Lambda$, we should expect corrections to Newton's law of
the form $(|\Lambda|r^2)^n$ for some positive constant $n$. Because of the very
small observed value of $\Lambda$, this correction will only be noticable at
very large distances. But the Newtonian potential is only a meaningful quantity
at short (i.e. non-cosmological) distances, so for our purpose the
approximation $\Lambda \approx 0$ should be perfectly valid. The same is also
true for a potentially non-vanishing $k$. From a pure mathematical point of
view, however, it would be interesting to consider an arbitrary value of
$\Lambda$. But in that case, one has to use the full deSitter propagator
instead of (\ref{eq:4dpropagator}).

Another way of looking at this problem, is to note that the effect of the fifth
dimension and the effect of $\Lambda$ and $k$ takes place at completely
different scales. The effect of the fifth dimension dominates at very short
distances (as we will see), where we can set $\Lambda = 0$ and $k = 0$, whereas
the effect of $\Lambda$ and $k$ dominates at very large distances, where we can
ignore the fifth dimension altogether and simply use 4-dimensional FRW space.
Trying to include both effects in the same calculation will only make things
unnecessarily complicated.

\subsection{The Newtonian potential}

Combining the above results (\ref{eq:4dpropagator}) and
(\ref{eq:gravitonvertex}) for the propagator and the vertex and summing over
$m$, we finally obtain the gravitational potential
\be
  V(\mathbf{k}) = \left.
    \sum_m |u(m,0)|^2 \frac{1}{\mu\MPl^2}
    \frac{T_1^{\mu\nu} P^{(m)}_{\mu\nu\alpha\beta} T_2^{\alpha\beta}}
      {|k^2 - m^2|}
  \right|_{k^0 \to 0}.
\ee
Taking both particles to be point particles at rest with masses $m_1$ and
$m_2$, the energy momentum tensors can be written $T_1^{\mu\nu}(\mathbf{x}) =
m_1 \delta(\mathbf{x}) u^\mu u^\nu \Rightarrow T_1^{\mu\nu}(\mathbf{k}) = m_1
u^\mu u^\nu = m_1 \delta^\mu_0 \delta^\nu_0$, and likewise for
$T_2^{\alpha\beta}$. We therefore see that only the 0000-component of the
polarization tensor contributes to the potential:
\be
  V(\mathbf{k}) =
    \frac{8\pi G m_1 m_2}{\mu} \sum_m |u(m,0)|^2
    \frac{P^{(m)}_{0000}(\mathbf{k})}{\mathbf{k}^2 + m^2} \, .
\ee
(Here we have written $\MPl^{-2} = 8\pi G$, where $G$ is the gravitational
constant.) Now, since $k_0 = 0$, the polarization tensor is easily found from
(\ref{eq:polarization_tensor1}) and (\ref{eq:polarization_tensor2}):
\be
  P^{(m)}_{0000}(\mathbf{k}) = \left\{ \begin{array}{lc}
    \ds \frac{1}{2} \, , & m = 0 \, , \vs \\
    \ds \frac{1}{2} + \frac{1}{6} = \frac{2}{3} \, , & m > 0 \, ,
  \end{array} \right.
\ee
with the result
\be
  V(\mathbf{k}) = \frac{8\pi G m_1 m_2}{\mu} \left\{
    \frac{1}{2} \frac{|u(0,0)|^2}{\mathbf{k}^2} +
    \frac{2}{3} \sum_{m>0} \frac{|u(m,0)|^2}{\mathbf{k}^2 + m^2}
  \right\},
\ee
or
\be
  V(r) = \frac{G m_1 m_2}{\mu r}
    \left\{ |u(0,0)|^2 + \frac{4}{3} \sum_{m>0} |u(m,0)|^2 e^{-mr} \right\}.
  \label{eq:gravpot1}
\ee
Thus, we see the importance of using the full tensor structure of the
propagator, instead of just the scalar part which is commonly used. Without the
tensor part we would miss the relative factor $4/3$ between the massless and
the massive modes.

In order to proceed further, we need the normalization constants $N_m$ and
$N_0$ in (\ref{eq:u_solution}) and (\ref{eq:u_solution_zero}), which in
principle are found by requiring
\ba
  \int_{-\infty}^\infty |u(0,z)|^2 dz \eq 1 \, , \nonumber \\
  \int_{-\infty}^\infty u(m,z)^* u(m',z) dz \eq \delta(m-m') \, .
\ea
However, the second integral as it stands is divergent for all $m$ and $m'$,
not just for $m=m'$, and therefore has to be regulated in some way. As
mentioned in \cite{RandallSundrum2}, this can be done by introducing a
regulator brane at some large but finite position $z_r$, and then taking the
limit $z_r \to \infty$ at the end. Since this point seems to be taken very
lightly in the literature, we will go through the calculations in detail here.

\subsection{Using a regulator brane}

With a second brane located at $y=y_r$, we still assume that the points $y$ and
$-y$ are equal, which means that the fifth dimension is now compactified on an
orbifold $S^1/Z_2$. The two branes thus represent the endpoints of the fifth
dimension. Einstein's equation in the bulk space and the boundary conditions on
the physical brane are unchanged by the presence of the regulator brane, and
the only way it manifests itself is as an additional set of boundary
conditions. (Several works have been done in the two brane scenario where both
branes are considered physical, see e.g. \cite{Brevik2}. In this paper we are
only using the second brane as a regulator, therefore ignoring its cosmological
implications.) From the energy momentum tensor $T^r_{MN} = \diag(\rho_r n^2,
p_r a^2 \gamma_{ij}, 0) \delta(y-y_r)$ on the regulator brane, we get
\ba
  \left.\frac{[a']}{a}\right|_{y=y_r} \eq -\tfrac{1}{3} M^{-3} \rho_r,
    \nonumber \\
  \left.\frac{[n']}{n}\right|_{y=y_r} \eq \tfrac{1}{3} M^{-3} (2\rho_r + 3p_r),
  \label{eq:boundary_regulator}
\ea
where $[a']_{y=y_r} = a'(y=y_r^+) - a'(y=y_r^-) = -2a'(y_r^-)$. Since $a(y)$ is
fixed by (\ref{eq:scalefactor_general}) or (\ref{eq:scalefactor_simple}), the
energy density $\rho_r$ on the regulator brane can not be chosen independently,
but is in fact directly related to the energy density $\rho_b$ on the physical
brane.

The condition (\ref{eq:boundary_regulator}) gives an additional term in the
potential for the wavefunction,
\be
  V(z) = \frac{15\mu^2}{4(1 + \mu|z|)^2} - 3\mu\delta(z) +
    \frac{3\mu}{1 + \mu z_r} \delta(z-z_r) \, ,
  \label{eq:potential_regulator}
\ee
and thus the new boundary condition
\be
  2u'(z_r) - 3A'(y_r)u(z_r) =
    2u'(z_r) + \frac{3\mu}{1 + \mu z_r} u(z_r) = 0 \, ,
  \label{eq:u_boundary_regulator}
\ee
in addition to (\ref{eq:u_boundary}). The zero mode (\ref{eq:u_solution_zero})
is trivially seen to satisfy this new condition. But the continuous spectrum of
massive modes is reduced to a discreet spectrum. Using (\ref{eq:u_solution})
together with well known identities for the Bessel functions,
(\ref{eq:u_boundary_regulator}) can be simplified to
\be
  \frac{Y_1(\tfrac{m}{\mu})}{J_1(\tfrac{m}{\mu})} =
    \frac{Y_1[\tfrac{m}{\mu}(1+\mu z_r)]}{J_1[\tfrac{m}{\mu}(1+\mu z_r)]}
  \stackrelraise{1mm}{z_r \to \infty}{\longrightarrow}
    \tan(m z_r - \tfrac{3\pi}{4}) \, ,
\ee
which means that the mass is approximately quantized in units of $\pi/z_r$:
\be
  m_n \simeq \frac{n\pi}{z_r} \, , \;\; n = 0, 1, 2, 3, \ldots
\ee
The transition from a sum over $m$ to a continuous integral in the limit $z_r
\to \infty$ is therefore given by
\be
  \sum_m f(m) = \sum_m f(m) \frac{z_r}{\pi} \Delta m \to
    \int_0^\infty f(m) \frac{z_r}{\pi} dm \, .
\ee

Next, we consider the normalization constant $N_m$ when using a regulator
brane, which is found from
\be
  \int_{-z_r}^{z_r} u_i(z) u_j(z) dz =
    2 \int_0^{z_r} u_i(z) u_j(z) dz = \delta_{ij} \, ,
\ee
since the eigenfunctions $u_i(z) \equiv u(m_i,z)$ in (\ref{eq:u_solution}) are
real. When $i \neq j$, we can use (\ref{eq:u_Schrodinger}) and write the
integrand as
\be
  u_i(z) u_j(z) = \frac{1}{m_i^2-m_j^2}
    \left[ u_i(z) u''_j(z) - u_j(z) u''_i(z) \right].
\ee
Performing a partial integration and using the boundary conditions at both
endpoints, the integral then vanishes identically in the case $i \neq j$. For
$i=j$ we get
\ba
  & & \hspace{-11mm} \frac{1}{2} =
    \int_0^{z_r} u^2(m,z) dz = \int_0^{z_r} N_m^2 (1+\mu z) \nonumber \\
  & & \hspace{-7mm} \times \! \left\{
    Y_2[\tfrac{m}{\mu}(1+\mu z)] -
    \frac{Y_1(\frac{m}{\mu})}{J_1(\frac{m}{\mu})} J_2[\tfrac{m}{\mu}(1+\mu z)]
  \right\}^2 \! dz \, .
\ea
In the limit $z_r \to \infty$ we can use the asymptotic expressions for $Y_2$
and $J_2$ for large arguments, with the result
\ba
  & & \hspace{-6mm}
    (z_r N_m^2)^{-1} \to \lim_{z_r \to \infty} \frac{1}{z_r} \int_0^{z_r}
    \frac{4\mu}{\pi m} \nonumber \\
  & & \hspace{2mm} \times \left\{
      \sin(mz-\tfrac{5\pi}{4}) -
      \frac{Y_1(\frac{m}{\mu})}{J_1(\frac{m}{\mu})} \cos(mz-\tfrac{5\pi}{4})
    \right\}^2 dz \nonumber \\
  & & \hspace{-3mm} = \frac{2\mu}{\pi m} \left[
      1 + \frac{Y_1^2(\frac{m}{\mu})}{J_1^2(\frac{m}{\mu})}
    \right],
\ea
or
\be
  N_m^2 = \frac{\pi m}{2\mu z_r} \left[
    1 + \frac{Y_1^2(\frac{m}{\mu})}{J_1^2(\frac{m}{\mu})}
  \right]^{-1}.
\ee
For the zero mode (\ref{eq:u_solution_zero}) we take the limit $z_r \to \infty$
directly, with the result
\be
  N_0^2 =
  \left[ \int_{-\infty}^\infty \frac{dz}{(1+\mu|z|)^3} \right]^{-1} = \mu \, .
\ee

Inserting everything into (\ref{eq:gravpot1}), we see that the factor $z_r$
from the integration measure and from the normalization constant cancel, as it
should. We then finally obtain the gravitational potential
\ba
  V(r) \eq \frac{G m_1 m_2}{\mu r}
    \left\{
      u^2(0,0) +
      \frac{4}{3} \int_0^\infty \!\! u^2(m,0) e^{-mr} \frac{z_r}{\pi} dm
    \right\} \nonumber \\
  \eq \frac{G m_1 m_2}{r} \left\{ 1 + \frac{2}{3\mu^2} \int_0^\infty m e^{-mr}
    \phantom{\frac{\left[J_1(\tfrac{m}{\mu})\right]^2}{J_1^2(\tfrac{m}{\mu})}}
    \right. \nonumber \\
  & & \hspace{3mm} \left. \times \frac{\left[
      J_1(\tfrac{m}{\mu}) Y_2(\tfrac{m}{\mu}) -
      Y_1(\tfrac{m}{\mu}) J_2(\tfrac{m}{\mu})
    \right]^2}{J_1^2(\tfrac{m}{\mu}) + Y_1^2(\tfrac{m}{\mu})}
    dm \right\} \nonumber \\
  \eq \frac{G m_1 m_2}{r} \left\{ 1 + \frac{8}{3\pi^2} \int_0^\infty
    \frac{e^{-mr}}{J_1^2(\tfrac{m}{\mu}) + Y_1^2(\tfrac{m}{\mu})} \frac{dm}{m}
  \right\}, \nonumber \\
  & & \label{eq:gravpot2}
\ea
where we have used the general identity $J_n(x) Y_{n+1}(x) - Y_n(x) J_{n+1}(x)
= -2/\pi x$ in the last line. The first term clearly gives 4-dimensional
gravity, and the integral over the massive modes gives the correction due to
the fifth dimension. Since the only parameter in the integral is $\mu r$ (after
substituting the dimensionless variable $n = m/\mu$), we see that the scale
where the correction becomes important is given by the curvature radius
$\mu^{-1}$ of the fifth dimension. For distances $r$ much larger than this, the
exponential factor ensures that 4-dimensional gravity is restored. The result
(\ref{eq:gravpot2}) is the same as that obtained by Chung \textit{et al.}
\cite{Chung}.

\begin{figure}[!h]
  \begin{center}
    \includegraphics[width=75mm]{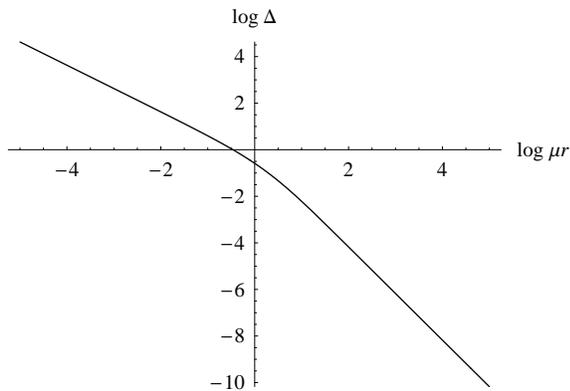}
  \end{center}
  \vspace{-7mm}
  \caption{The figure shows the relative correction $\Delta$ in
    (\ref{eq:gravpot2}), where $V(r) = V_0(r) (1 + \Delta)$, as obtained by a
    numerical integration using {\Mathematica}. The leading order terms $\Delta
    \simeq 4/(3\pi\mu r)$ (short distances) and $\Delta \simeq 2/(3\mu^2 r^2)$
    (large distances) are clearly seen from the figure, and the transition
    between the two regions occurs around $\mu r \sim 1$.}
  \label{fig:gravpot}
\end{figure}

The integral in (\ref{eq:gravpot2}) must in general be done numerically, which
is straightforward using {\Mathematica}, and the result is shown in figure
\ref{fig:gravpot}. It is also possible to expand the result in powers of $\mu
r$ for the two regions $\mu r \ll 1$ and $\mu r \gg 1$. Writing the full
potential as $V(r) = V_0(r) (1 + \Delta)$, the relative correction $\Delta$ has
the series expansion
\be
  \Delta = \left\{ \begin{array}{lc}
    \ds \frac{4}{3\pi \mu r} - \frac{1}{3} -
      \frac{1}{2\pi} \mu r \ln \mu r & \vs \\
    \ds \hspace{5mm} + \, 0.089237810 \mu r + {\cal O}(\mu^2 r^2) \, ,
    & \mu r \ll 1, \vs \\
    \ds \frac{2}{3\mu^2 r^2} - \frac{4\ln \mu r}{\mu^4 r^4} +
      \frac{16 - 12\ln 2}{3\mu^4 r^4} & \vs \\
    \ds \hspace{5mm} + \,
      {\cal O} \!\left[ \frac{(\ln \mu r)^2}{\mu^6 r^6} \right] \! ,
    & \mu r \gg 1.
  \end{array} \right.
\ee
A detailed derivation of this result is included in appendix
\ref{cha:gravpot_series}. Note in particular the leading order term $\Delta
\simeq 4/(3\pi\mu r)$ for short distances. When $\mu r \ll 1$, $\Delta \gg 1$
and the entire gravitational potential behaves like $V \sim 1/r^2$. Gravity is
therefore 5-dimensional at short distances. This should not come as a surprise,
since when the distance $r$ is small compared to the curvature radius
$\mu^{-1}$ of the fifth dimension, spacetime looks almost flat and we should
get the same result as with 5-dimensional Minkowski space.

\section{Summary}

In this paper, we have gone through the derivation of the Newtonian
gravitational potential in the RS model, from start to finish. We have
concentrated on the critical case ($\Lambda = 0$ and $k = 0$), and have
obtained the result (for large distances) $V \sim \frac{1}{r}(1 +
\frac{2}{3\mu^2 r^2} + \ldots)$ from the graviton propagator, by including its
full tensor structure. This is exactly the same result as what is found from a
matter source on the brane when the brane-bending effect is included. As for
the non-critical case, since both $\Lambda$ and $k/a_0^2$ are very small (from
observational constraints), corrections due to $\Lambda$ and $k$ will be
noticable only at cosmological distances, and the correction due to the fifth
dimension must therefore be the same as for the critical case
(\ref{eq:gravpot2}).

Throughout this paper, we have assumed that the dark radiation term vanishes
($U=0$), and we have ignored contributions from matter and radiation on the
brane by setting $\rho = \rho_m + \rho_r = 0$. In a realistic model, however,
these terms have to be included. Because of the degeneracy between dark
radiation and ordinary radiation, the two of them can not be distinguished by
studying cosmological evolution alone. But since no observation has been made
that requires the presence of dark radiation, we can probable assume that it is
small, i.e. that $|U| \lesssim \rho_r a_0^4 / \MPl^2$.

It is interesting to note, however, that the scale factor $a(t,y)$ may change
dramatically far from the brane when including even a tiny value of $\rho$ or
$U$, which is seen from the following example: For $\Lambda_B < 0$ and $\Lambda
= 0$ we assume that all parameters of the model ($M$, $\mu$ and $\lambda$) are
of the order of the Planck scale, which means that we ignore the $\rho^2$ term
in (\ref{eq:Friedmann1_simple}). Combining this with
(\ref{eq:scalefactor_general}), the scale factor reduces to
\be
  \frac{a^2(t,y)}{a_0^2(t)} = e^{-2\mu|y|} -
    \frac{\rho \left( 1 - e^{-2\mu|y|} \right)}{6\mu^2 \MPl^2} -
    \frac{U \sinh^2 \mu y}{\mu^2 a_0^4} \, .
\ee
Thus, both $\rho$ and $U$ introduce a horizon at a finite distance from the
brane, unless $U$ is negative, in which case the scale factor reaches a minimum
and then diverges as $y \to \infty$. The same two effects arise when $\Lambda
> 0$ and $\Lambda < 0$, respectively. However, as shown by Karch and Randall
\cite{Karch_Randall}, gravity can still be localized in these cases, since
localization only depends on the scale factor close to the brane. We should
therefore expect that matter and (dark) radiation only play a significant role
in the very early universe, when $\rho$ and $U/a_0^4$ are of the order of the
Planck scale.

\vspace{4mm}\textbf{Acknowledgement:} This work has been supported by grant no.
NFR 153577/432 from the Research Council of Norway.

\appendix

\section{Series expansion of $\Delta$}
\label{cha:gravpot_series}

We are considering the integral
\be
  \Delta = \frac{8}{3\pi^2} \int_0^\infty
    \frac{e^{-\mu rn}}{J_1^2(n) + Y_1^2(n)} \frac{dn}{n} \equiv
    \int_0^\infty f(n) e^{-\mu rn} dn \, ,
\ee
where we are using the dimensionless integration variable $n = m/\mu$. In the
following, we will need the series expansion of $f(n)$ for small and large
values of $n$, which is found using {\Mathematica}:
\ba
  \hspace{-8mm} f(n \ll 1) \eq \frac{2}{3}n +
    \frac{2}{3} \left( -\frac{1}{2} + \gamma - \ln 2 + \ln n \right) n^3
    \nonumber \\
  & & \hspace{-4mm} + \left[ -0.107720074417 + \frac{1}{2}(\ln n)^2 \right.
    \nonumber \\
  & & \left. \: +
      \left( \gamma - \ln 2 - \frac{7}{12} \right) \ln n
    \right] n^5 + \ldots,
  \label{eq:f_expansion1}
\ea \vspace{-5mm}
\be
  f(n \gg 1) = \frac{4}{3\pi} - \frac{1}{2\pi n^2} + \frac{21}{32\pi n^4} -
    \frac{633}{256\pi n^6} + \ldots
  \label{eq:f_expansion2}
\ee

\subsection{Short distances, $\mu r \ll 1$}

In the limit of short distances, the exponential factor $e^{-\mu rn}$ becomes
important when $n$ is large and we can use (\ref{eq:f_expansion2}). However,
this series can not be integrated term by term from 0 to $\infty$, since all
terms (except the first one) diverges in the limit $n \to 0$. We therefore
split the integral into two parts, $\int_0^\infty = \int_0^N + \int_N^\infty$,
where $N$ is assumed to be large. In the first part we expand $e^{-\mu rn}$,
and the second part is integrated term by term, using the general formula
\ba
  \int_N^\infty \frac{e^{-\mu rn}}{n^k} dn \eq
    e^{-\mu rN} \sum_{i=0}^{k-2}
      \frac{(k-2-i)!}{(k-1)!} \frac{(-\mu r)^i}{N^{k-1-i}} \nonumber \\
  & & + \frac{(-\mu r)^{k-1}}{(k-1)!} \Gamma(0, \mu rN) \, .
\ea
Here $\Gamma(a,z)$ is the incomplete gamma function $\Gamma(a,z) =
\int_z^\infty t^{a-1} e^{-t} dt$. Writing the expansion of $f(n)$ as $f(n) =
\sum_{k=0}^\infty C_{2k} n^{-2k}$, we then obtain
\ba
  \Delta \eq
    \sum_{k=0}^\infty \frac{(-\mu r)^k}{k!} \int_0^N n^k f(n) dn +
    \frac{4e^{-\mu rN}}{3\pi\mu r} \nonumber \\
  & & \hspace{-2mm} + \sum_{k=1}^\infty C_{2k} \left[
    e^{-\mu rN} \sum_{i=0}^{2k-2}
      \frac{(2k-2-i)!}{(2k-1)!} \frac{(-\mu r)^i}{N^{2k-1-i}}
  \right. \nonumber \\
  & & \left. \hspace{7.4mm} \phantom{\sum_{i=0}^{2k-2}} -
    \frac{(\mu r)^{2k-1}}{(2k-1)!} \Gamma(0, \mu rN)
  \right].
  \label{eq:gravpot_expansion1}
\ea
This expression can now be expanded in powers of $\mu r$, using
\be
  \Gamma(0, \mu rN) = -\gamma - \ln \mu r - \ln N -
    \sum_{k=1}^\infty \frac{(-\mu rN)^k}{k! \cdot k} \, ,
  \label{eq:Gamma_expansion}
\ee
and we can then extract and calculate the coefficient in front of each power of
$\mu r$ in turn. From (\ref{eq:Gamma_expansion}), we see that we also get
several logarithmic terms. For each power of $\mu r$, we take the limit $N \to
\infty$, therefore ignoring all terms with a negative power of $N$. Using the
notation
\be
  \Delta = \sum_{k=-1}^\infty A_k (\mu r)^k + \mathrm{logarithmic \; terms} ,
\ee
we first find rather trivially that $A_{-1} = 4/(3\pi)$, and then, using
numerical integration:
\begin{widetext} \vspace{-4mm}
\ba
  A_0 \eq
    \lim_{N \to \infty} \left[ \int_0^N f(n) dn - \frac{4N}{3\pi} \right] =
    \int_0^\infty \left[ f(n) - \frac{4}{3\pi} \right] dn = -\frac{1}{3} \, ,
    \nonumber \\
  A_1 \eq
    \lim_{N \to \infty} \left[
      -\int_0^N n f(n) dn + \frac{2N^2}{3\pi} - \frac{1}{2\pi} \ln N
    \right] + \frac{1-\gamma}{2\pi} \nonumber \\
  \eq -\int_0^1 n \left[ f(n) - \frac{4}{3\pi} \right] dn -
    \int_1^\infty n\left[f(n)-\frac{4}{3\pi}+\frac{1}{2\pi n^2}\right] dn +
    \frac{1-\gamma}{2\pi} = 0.08923780957038536 \, , \nonumber \\
  A_2 \eq
    \lim_{N \to \infty} \left[
      \frac{1}{2} \int_0^N n^2 f(n) dn - \frac{2N^3}{9\pi} + \frac{N}{4\pi}
    \right] =
    \frac{1}{2} \int_0^\infty n^2 \left[
      f(n) - \frac{4}{3\pi} + \frac{1}{2\pi n^2}
    \right] dn = \frac{1}{8} \, , \nonumber \\
  A_3 \eq
    \lim_{N \to \infty} \left[
      -\frac{1}{3!} \int_0^N n^3 f(n) dn + \frac{N^4}{18\pi} -
      \frac{N^2}{24\pi} + \frac{7}{64\pi} \ln N
    \right] + \frac{42\gamma - 77}{384\pi} \nonumber \\
  \eq -\frac{1}{6} \int_0^1 n^3 \left[
      f(n) - \frac{4}{3\pi} + \frac{1}{2\pi n^2} \right] dn -
    \frac{1}{6} \int_1^\infty n^3 \left[
      f(n) - \frac{4}{3\pi} + \frac{1}{2\pi n^2} - \frac{21}{32\pi n^4}
    \right] dn + \frac{42\gamma - 77}{384\pi} \nonumber \\
  \eq -0.02955870986828890 \, , \phantom{M^P} \nonumber \\
  A_4 \eq
    \lim_{N \to \infty} \left[
      \frac{1}{4!} \int_0^N n^4 f(n) dn - \frac{N^5}{90\pi} +
      \frac{N^3}{144\pi} - \frac{7N}{256\pi}
    \right] \nonumber \\
  \eq \frac{1}{24} \int_0^\infty n^4 \left[
      f(n) - \frac{4}{3\pi} + \frac{1}{2\pi n^2} - \frac{21}{32\pi n^4}
    \right] dn = -\frac{3}{128} \, .
\ea
\end{widetext}
Note in particular how the first terms in the expansion of $f(n)$ act as
counterterms, making all the integrals converge, and the numerical calculation
therefore straightforward using {\Mathematica}. The coefficient in front of the
logarithmic terms is easily found from (\ref{eq:gravpot_expansion1}) and
(\ref{eq:Gamma_expansion}):
\be
  \sum_{k=1}^\infty \frac{C_{2k}}{(2k-1)!} (\mu r)^{2k-1} =
    -\frac{1}{2\pi} \mu r + \frac{7}{64\pi} (\mu r)^3 + \ldots \hspace{2mm}
\ee
The complete series expansion of $\Delta$ is therefore obtained as
\ba
  \Delta \eq \frac{4}{3\pi \mu r} - \frac{1}{3} + 0.089237810\mu r -
    \frac{1}{2\pi}\mu r \ln \mu r \nonumber \\
  & & \hspace{-1mm} + \, \frac{1}{8}\mu^2 r^2 - 0.029558710\mu^3 r^3 +
    \frac{7}{64\pi}\mu^3 r^3 \ln \mu r \nonumber \\
  & & \hspace{-1mm} - \, \frac{3}{128}\mu^4 r^4 + {\cal O}(\mu^5 r^5) \, .
\ea

\subsection{Large distances, $\mu r \gg 1$}

In the limit of large distances, the series expansion of $\Delta$ is obtained a
lot easier than in the previous section, since we can now integrate the
expansion of $f(n)$ in (\ref{eq:f_expansion1}) for small $n$ directly from 0 to
$\infty$. The result is therefore
\ba
  \Delta \eq \frac{2}{3\mu^2 r^2} + \frac{16-12\ln 2}{3\mu^4 r^4} -
    \frac{4\ln \mu r}{\mu^4 r^4} + \frac{29.4398446730}{\mu^6 r^6}
    \nonumber \\
  & & \hspace{-2.2mm} +\: \frac{(120\ln 2 - 204)\ln \mu r}{\mu^6 r^6} +
    \frac{60(\ln \mu r)^2}{\mu^6 r^6} + {\cal O}(1 / \mu^8 r^8) \, .
    \nonumber \\
  & &
\ea
The error that follows from using the expansion (\ref{eq:f_expansion1}) all the
way to $n = \infty$ is of exactly the same order as what we would get from the
next term in the expansion, which is easily shown: If we denote the first terms
in the expansion by $h(n)$, such that $f(n) = h(n) + {\cal O}(n^7)$ for small
$n$, the relative correction to the gravitational potential can be written
$\Delta = \int_0^\infty f(n) e^{-\mu rn} dn = \int_0^\infty h(n) e^{-\mu rn} dn
+ \delta$, where the error $\delta$ is bounded by
\be
  |\delta| \leq \int_0^\infty e^{-\mu rn}
    \left| \raisebox{0mm}[2mm][2mm]{$\ds f(n)-h(n)$} \right| dn \, .
\ee
It turns out that $f(n) - h(n) = \frac{1}{3} n^7 (\ln n)^3 + \!$ higher order
terms. A rough estimate $E(n)$ of the difference $f(n)-h(n)$ which satisfies
$E(n) > |f(n)-h(n)|$ for all $n$ can therefore be chosen as $E(n) = n^7 \left[
1 - (\ln n)^3 e^{-n} \right]$. This means that
\be
  |\delta| < \int_0^\infty \!\! e^{-\mu rn} E(n) dn
  \sim \frac{(\ln \mu r)^3}{\mu^8 r^8} +
    \mathrm{higher \; order \; terms} ,
\ee
i.e. precisely the same leading order term as what is obtained from the first
term not included in (\ref{eq:f_expansion1}).

\end{document}